\documentclass[12pt]{article}
\usepackage{amssymb,graphicx}
\usepackage{epstopdf}
\usepackage{amsmath,amsfonts}
\usepackage{epsfig}
\usepackage[titletoc,title]{appendix}
\def\const{\mbox{const}}

\newcommand{\be}{\begin{equation}}
\newcommand{\ee}{\end{equation}}
\newcommand{\bea}{\begin{eqnarray}}
\newcommand{\eea}{\end{eqnarray}}
\newcommand{\bg}{\begin{gather}}
\newcommand{\eg}{\end{gather}}
\newcommand{\bseq}{\begin{subequations}}
	\newcommand{\eseq}{\end{subequations}}

\usepackage[english]{babel}
\usepackage{amsmath,amssymb} 
\usepackage{ mathrsfs }
\newcommand{\RNumb}[1]{\uppercase\expandafter{\romannumeral #1\relax}}
\numberwithin{equation}{section}

\begin{document}
	\begin{flushright}
	\end{flushright}
	\vspace{10pt}
	\begin{center}
		{\LARGE \bf Galileon-like vector fields.} \\
		\vspace{20pt}
	P. K. Petrov$^{a,b}$\\
	\vspace{15pt}
	$^a$\textit{
Department of Particle Physics and Cosmology, Faculty of Physics,
M. V. Lomonosov Moscow State University, Vorobyovy Gory, 1-2, Moscow,
119991, Russia
	}\\
	\vspace{5pt}
	$^b$\textit{
		Institute for Nuclear Research of
		the Russian Academy of Sciences,\\  60th October Anniversary
		Prospect, 7a, 117312 Moscow, Russia}\\
	\vspace{5pt}
	\end{center}
	\vspace{5pt}
	
	\begin{abstract}
We construct simple Lagrangians of vector fields which involve second derivatives, but nevertheless lead to second order field equations. These vector fields are, therefore, analogs of  generalized Galileons. Our construction is given first in Minkowski space, and then generalizied to include dynamical gravity. We present examples of backgrounds that are stable and ghost-free despite the absence of gauge invariance. Some of these backgrounds violate the Null Energy Condition. 
\end{abstract}
                  \section{Introduction and summary.}

Scalar theories with Lagrangians involving second derivatives, which nevertheless lead to second order field equations, attract considerable interest. These are theories of generalized Galileons \cite{1} whose versions with dynamical gravity are Horndeski theories \cite{2,3}.
From the cosmological viewpoint, these theories are particularly interesting because they are capable of violating the Null Energy Condition (NEC) in a healthy way (for a review see \cite{4}).
It is natural to try to generalize these theories to fields other than scalar. If one insists on gauge  invariance, then no generalization is possible in four dimensions \cite{5} while in higher dimensions one arrives at a theory of p-form Galileons \cite{6}, \cite{7}.
Giving up gauge invariance is dangerous but may not be fatal. Indeed, there are vector theories (with  Lagrangians  involving first derivatives only) which are not gauge invariant but, nevertheless, stable. One class of such theories is the generalized Proca theories, or vector Galileons \cite{8,9}. Theories of another class \cite{10} are stable in non-trivial backgrounds.
An interesting property of the latter is that they also may violate the NEC in a healthy way \cite{11}. 

In this paper we also consider vector field and give up gauge invariance. Our purpose is to construct the simplest vector-field Lagrangians involving second derivatives and yet giving rise to second order field equations. We do this first in Minkowski space and find that there are at least three fairly large classes of theories that have the desired property. We then switch on the dynamical gravity and observe that all field equations remain second order for theories belonging to  two of these classes. We consider one of these classes further and give an example of  vector background in Minkowski space that violates the NEC.
Then  we derive the conditions for stability (absence of ghosts and gradient instabilities) about this background in Minkowski space and find the range of parametrs where the NEC-violating background is stable.
Thus, theories we consider may be viewed as vector analogs of the generalized Galileons.

This paper is organized as follows.
In  Section 2 we construct non gauge-invariant second derivative Lagrangians with second order field equations for vector fields in Minkowski space. 
In  Section 3 we turn on dynamical gravity and show that all equations of motion remain second order in theories belonging to two of the classes found in Section 2.
In Section 4 we give an example of non-trivial homogeneous vector field background in Minkowski space that violates the NEC and derive the conditions for stability and  for the absence of  superluminal propogation of perturbations. Finally, we determine the range of parameters, in which the background is stable and violates the NEC in Minkowski space.  \vspace{0.3 cm}

\section{Second-derivative Lagrangians in Minkowski space.} 
Let us construct a non gauge-invariant theory for vector field in Minkowski space which has the Lagrangian satisfying the following requirements:\\

\begin{enumerate}
\item The Lagrangian $\mathscr{L}$ has second derivatives, along with first derivatives and the field itself. 
\item Field equations obtained from this Lagrangian  have derivatives of at most second order.  
\item  The Lagrangian cannot be reduced by integration by parts to the Lagrangian involving first derivatives only.
\end{enumerate}
We are going to construct the simplest theories, for which:
\begin{enumerate}
\item[4.] The Lagrangian is linear in the second derivatives:
\begin{equation}
\mathscr{L}=S^{\mu\nu\rho}(A_{\lambda} ;A_{\tau;\xi})A_{\rho;\mu\nu}+L(A_{\tau}, A_{\lambda;\xi})
\label{1.1}
\end{equation}
\end{enumerate}
It is convenient to think of $S^{\mu\nu\rho}$ as a sum
\begin{equation}
S^{\mu\nu\rho} = \frac{1}{2}(K^{\mu\nu\rho}+K^{\nu\mu\rho}), 
\label{1.4}
\end{equation} 
where $K^{\mu\nu\rho}$ does not have to be symmetric in $\mu,$ $\nu$.\\

Our last simplifying assumption is
\begin{enumerate}
\item[5.]The function $K^{\mu\nu\rho}$ in (\ref{1.4}) is a monomial in variables $A_{\mu},$ $A_{\nu;\tau}$
which does not involve the totally antisymmetric tenzor:
\begin{equation}
K^{\mu_{\alpha}\mu_{\beta}\mu_{\gamma}}=\const \cdot\eta^{\mu_{\sigma(1)}\mu_{\sigma(2)}}...\eta^{\mu_{\sigma(n+2m+2)}\mu_{\sigma(n+2m+3)}}A_{\mu_{1}}...A_{\mu_{n}}A_{\mu_{n+1};\mu_{n+2}}...A_{\mu_{n+2m-1};\mu_{n+2m}},
\label{1.95}
\end{equation}
where $n$ is odd, $\sigma$ denotes a permutation of $(n+2m+3)$ indices, and $\mu_{\alpha}, $ $\mu_{\beta},$ $\mu_{\gamma}$ are non-convoluted indeces,  $(\mu_{\alpha}, \mu_{\beta}, \mu_{\gamma})=(\mu_{\sigma^{-1}(n+2m+1)}, \mu_{{\sigma^{-1}(n+2m+2)}}, \mu_{{\sigma^{-1}(n+2m+3)}}).$
\end{enumerate}
The Euler-Lagrange equations for a theory with this Lagrangian have the following form:
\begin{equation}
\frac{\partial \mathscr{L}}{\partial A_{\rho}}-\partial_{\mu}\frac{\partial \mathscr{L}}{\partial A_{\rho;\mu}}+\partial_{\mu}\partial_{\nu}\frac{\partial \mathscr{L}}{\partial A_{\rho;\mu\nu}}=0,
\label{1.0}
\end{equation} 
where $A_{\rho;\mu}=\partial_{\mu}A_{\rho},$ $A_{\rho;\mu\nu}=\partial_{\mu}\partial_{\nu}A_{\rho}.$ The third order terms in  eq. (\ref{1.0}) for the Lagrangian (\ref{1.1}) read:
\begin{equation*}
 \Big(\frac{\partial S^{\mu\nu\rho}}{\partial A_{\tau;\lambda}}-\frac{\partial S^{\mu\nu\tau}}{\partial A_{\rho;\lambda}} \Big)A_{\tau;\lambda\mu\nu}
\label{1.8}
\end{equation*}
Thus, to have second-order field equations, we require that
\begin{equation}
\frac{\partial S^{\mu\nu\rho}}{\partial A_{\tau;\lambda}}-\frac{\partial S^{\mu\nu\tau}}{\partial A_{\rho;\lambda}} =0.
\label{1.9}
\end{equation}

In accordance with (\ref{1.95}), the indices $\mu,$ $\nu,$ $\rho$ in the function $K^{\mu\nu\rho}$ come from the metrics or vector field or derivative of vector field.
The last index  in  $K^{\mu\nu\rho}$   plays a different role in eq. (\ref{1.9}) than the other  indices, and so it is convenient to classify  functions $K^{\mu\nu\rho}$ according to the  "origin" of the index $\rho$. In this way we arrive at four possibilities (other options give the same $S^{\mu\nu\rho}$ in (\ref{1.4})):
\begin{itemize}
\item[\RNumb{1}.] $K^{\mu\nu\rho}=L^{\mu\nu}_{\ \ \ \varkappa}(A_{\sigma}, A_{\tau;\lambda})A^{\varkappa;\rho}$
\item[\RNumb{2}.]$K^{\mu\nu\rho}=f^{\mu}(A_{\sigma}, A_{\tau;\lambda})\eta^{\nu\rho}$ 
\item[\RNumb{3}.] $K^{\mu\nu\rho}=B^{\mu\nu}(A_{\sigma}, A_{\tau;\lambda})A^{\rho}$
\item[\RNumb{4}.] $K^{\mu\nu\rho}=\widetilde{L}^{\mu\nu}_{\ \ \ \varkappa}(A_{\sigma}, A_{\tau;\lambda})A^{\rho;\varkappa}$,	
\end{itemize}
where functions $L^{\mu\nu}_{\ \ \alpha},$ $\widetilde{L}^{\mu\nu}_{\ \ \ \varkappa},$ $B^{\mu\nu}$ and $f^{\mu}$
are again monomials in two variables $A_{\sigma},$  $A_{\tau;\lambda}$ that do not involve totally antisymmetric tensor. Furthermore, it is convenient to classify the functions $\widetilde{L}^{\mu\nu}_{\ \ \ \varkappa}$ according to the "origin" of index $\varkappa$:
\begin{itemize}
\item[\RNumb{4.1}.]$\widetilde{L}^{\mu\nu}_{\ \ \ \varkappa}=T^{\mu}(A_{\sigma}, A_{\tau;\lambda})\delta_{\varkappa}^{\nu}$; \ \ \  
$K^{\mu\nu\rho}=T^{\mu}(A_{\sigma}, A_{\tau;\lambda})A^{\rho;\nu}$
\item[\RNumb{4.2}.] $\widetilde{L}^{\mu\nu}_{\ \ \ \varkappa}=X^{\mu\nu}_{\ \ \ \alpha}(A_{\sigma}, A_{\tau;\lambda})A^{\alpha;}_{\ \ \varkappa}$; \ \ \ 
    $K^{\mu\nu\rho}=X^{\mu\nu}_{\ \ \ \alpha}(A_{\sigma}, A_{\tau;\lambda})A^{\alpha;}_{\ \ \varkappa}A^{\rho;\varkappa}$
\item[\RNumb{4.3}.] $\widetilde{L}^{\mu\nu}_{\ \ \ \varkappa}=Z^{\mu\nu}_{\ \ \ \alpha}(A_{\sigma}, A_{\tau;\lambda})A^{\ \ \alpha}_{\varkappa;};$ \ \ \ 
 $K^{\mu\nu\rho}=Z^{\mu\nu}_{\ \ \ \ \alpha}(A_{\sigma}, A_{\tau;\lambda})A^{\ \ \alpha}_{\varkappa;}A^{\rho;\varkappa}$
\item[\RNumb{4.4}.] $\widetilde{L}^{\mu\nu}_{\ \ \ \varkappa}=V^{\mu\nu}(A_{\sigma}, A_{\tau;\lambda})A_{\varkappa};$
\ \ \ $K^{\mu\nu\rho}=V^{\mu\nu}(A_{\sigma}, A_{\tau;\lambda})A_{\varkappa}A^{\rho;\varkappa}.$
\end{itemize}
Making use of this classification, we analyse eq. (\ref{1.9}) in Appendix A. We find that there are three independent Lagrangians which satisfy above requirements 1 -- 5, namely
\begin{equation} 
\mathscr{L}_{1}=(F)^{l_{1}}(D)^{n_{1}}(B)^{k_{1}}\eta^{\mu\nu}A^{\rho}A_{\rho;\mu\nu},
\label{1.96}
\end{equation}\\
\begin{equation}
\mathscr{L}_{2}=(F)^{l_{2}}(D)^{n_{2}}(B)^{k_{2}}A_{\sigma}A_{\tau}A^{\sigma;\mu}A^{\tau;\nu}A^{\rho}A_{\rho;\mu\nu},
\label{1.97}
\end{equation}\\
\begin{equation}
\mathscr{L}_{3}=(F)^{l_{3}}(C)^{n_{3}}\eta^{\mu\nu}A^{\rho;\sigma}A_{\sigma}A_{\rho;\mu\nu},
\label{1.98}
\end{equation}\\
where $k_{i},$ $l_{i},$ $n_{i}$ are non-negative integers, and 
\begin{equation}F=A_{\mu}A^{\mu}, \label{1l}\end{equation}
\begin{equation}D=A^{\nu}A^{\lambda}A_{\nu; \lambda}, \label{2l}\end{equation}
\begin{equation}B=A_{\mu}A^{\nu}A^{\mu;\lambda}A_{\nu;\lambda}, \label{3l}\end{equation}
\begin{equation}C=A^{\mu;\tau}A_{\tau}A^{\rho}A_{\mu;\rho}. \label{4l}\end{equation}
The Lagrangians (\ref{1.96})  and (\ref{1.97}) have the structure corresponding to the case \RNumb{3} above for function $K^{\mu\nu\rho},$ while the Lagrangian (\ref{1.98}) corresponds to \RNumb{4.4}.

The Lagrangians (\ref{1.96}) -- (\ref{1.98}) contain  second derivatives, provided that
\begin{equation}
k_{1}\not=0 \ \ \mbox{and/or} \ \ n_{1}>1,
\end{equation}
\begin{equation}
k_{2}\not =0 \ \ \mbox{and/or} \ \ n_{2} \not=0,
\end{equation}
\begin{equation}
n_{3} \not= 0,
\end{equation}
respectively. Lagrangians (\ref{1.96}) and (\ref{1.97}) are independent  when
\begin{equation}
n_{1}>1.
\label{f}
\end{equation}\\
Straightforward generalizations of (\ref{1.96}) -- (\ref{1.98}) are  
\begin{equation} 
\begin{aligned}
&\mathscr{L}_{1}=f^{(1)}(B,D,F)\eta^{\mu\nu}A^{\rho}A_{\rho;\mu\nu}, \ \ \ 
&f^{(1)}_{B}\not=0 \ \  \mbox{and/or} \ \ f^{(1)}_{DD}\not=0,
\label{L1}
\end{aligned}
\end{equation}\\
\begin{equation}
\begin{aligned}
&\mathscr{L}_{2}=f^{(2)}(B,D,F)A_{\varkappa}A_{\tau}A^{\varkappa;\mu}A^{\tau;\nu}A^{\rho}A_{\rho;\mu\nu},\ \ \
&f^{(2)}_{B}\not=0 \  \mbox{and/or} \ \ f^{(2)}_{D}\not=0,
\label{L2}
\end{aligned}
\end{equation}\\ 
\begin{equation}
\begin{aligned}
&\mathscr{L}_{3}=f^{(3)}(C,F)\eta^{\mu\nu}A^{\rho;\lambda}A_{\lambda}A_{\rho;\mu\nu},\ \ \
&f^{(3)}_{C}\not=0 ,
\label{L3}
\end{aligned}
\end{equation}\\
where $f^{(1)},$ $f^{(2)}$ and $f^{(3)}$ are arbitary functions of their arguments, and $f_{B}=\frac{\partial f}{\partial B},$ $f_{DD}=\frac{\partial^{2}f}{\partial D^{2}},$ etc.

It is worth pointing out that there may exist linear combinations of Lagrangians whose structure is different from (\ref{L1}) -- (\ref{L3}), but which nevertheless lead to second order field equations due to cancellations between different terms. One of the examples is
$$
\mathscr{L}=\Big(\frac{1}{2}A^{\rho}A^{\mu;\nu}A_{\nu}A_{\mu;\lambda}A^{\lambda}+A^{\rho;\tau}A_{\tau}A_{\mu;\nu}A^{\mu}A^{\nu}\Big)\square A_{\rho}.
$$
We do not consider this fairly cotrived possibility in this paper.

\section{Turning on gravity.}
In the previous section we constructed three non gauge-invariant vector-field Lagrangians involving second derivatives and yet giving rise to second order and/or lower field equations in Minkowski space, eqs. (\ref{L1}) -- (\ref{L3}). Our purpose here to figure out which of these Lagrangians lead to the second order or lower equations of motion and energy-momentum tensor.

Let us consider the Lagrangian (\ref{L1}). One assumes minimal coupling to gravity, then  $-\sqrt{-g}T^{\rho\sigma}\delta g_{\rho\sigma}$ for this theory reads
\begin{equation}
\begin{aligned}
&-\sqrt{-g}T^{\rho\sigma}\delta g_{\rho\sigma}=2\delta ( \sqrt{-g} \mathscr{L}_{(1)})=\delta(\sqrt{-g}f^{(1)}(B,D,F)\square F)+... \\&\Rightarrow \sqrt{-g}f^{(1)}_{B}g^{\mu\nu}((\partial_{\mu}\partial_{\nu}F)\delta B+(\partial_{\mu}\partial_{\nu}B)\delta F)\\&+\sqrt{-g}f^{(1)}_{D}g^{\mu\nu}((\partial_{\mu}\partial_{\nu}F)\delta D+(\partial_{\mu}\partial_{\nu}D)\delta F)+...,
\label{2.1}
\end{aligned}
\end{equation}
where omitted terms do not contain third derivatives and arrow denotes integration by parts and $\delta B=\frac{\delta B}{\delta g_{\rho\sigma}}\delta g_{\rho\sigma},$  etc.
It is convient to represent eq. (\ref{2.1}) in the following form:
$$
-\sqrt{-g}T^{\rho\sigma}\delta g_{\rho\sigma}\Rightarrow I_{1}+I_{2}+...,
$$
where  
$$
I_{1}=\sqrt{-g}f^{(1)}_{B}g^{\mu\nu}((\partial_{\mu}\partial_{\nu}F)\delta B+(\partial_{\mu}\partial_{\nu}B)\delta F),	
$$
$$
I_{2}=\sqrt{-g}f^{(1)}_{D}g^{\mu\nu}((\partial_{\mu}\partial_{\nu}F)\delta D+(\partial_{\mu}\partial_{\nu}D)\delta F).	
$$
We see that $T^{\mu\nu}$ does not  contain third order derivatives of vector field and/or metric. Indeed, using the fact that $B=\frac{F_{;\mu}F^{;\mu}}{4},$ we obtain that $I_{1}$ is second order or lower
\begin{equation*}
\begin{aligned}
I_{1}&\Rightarrow \frac{\sqrt{-g}}{2}f^{(1)}_{B}(-(\partial^{\tau}F) (\partial_{\tau}\partial_{\mu}\partial_{\nu}F) \delta F+(\partial^{\tau}F) (\partial_{\mu}\partial_{\nu}\partial_{\tau}F) \delta F)+...=0+...
\label{2.3}
\end{aligned}
\end{equation*}
$I_{2}$ does not  contain third order derivatives too:
\begin{equation*}
\begin{aligned}
I_{2}&=f^{(1)}_{D}\sqrt{-g}A^{\lambda}A_{\varkappa}A^{\nu}A^{\rho}A^{\mu}g^{\sigma\alpha}((\partial_{\sigma}\partial_{\alpha} g_{\mu\nu})\delta\Gamma^{\varkappa}_{\rho\lambda}+(\partial_{\sigma}\partial_{\alpha} \Gamma^{\varkappa}_{\rho\lambda})\delta g_{\mu\nu})\\&-f^{(1)}_{D}\sqrt{-g}A^{\lambda}A^{\rho}g^{\sigma\alpha}
(2A^{\mu}(\partial_{\sigma}\partial_{\alpha}A_{\mu})A_{\varkappa}\delta\Gamma^{\varkappa}_{\rho\lambda}+
(\partial_{\sigma}\partial_{\alpha}\partial_{\lambda}A_{\rho})A^{\mu}A^{\nu}\delta g_{\mu\nu})\\&+...
\\&\Rightarrow
\frac{1}{2}f^{(1)}_{D}\sqrt{-g}A^{\lambda}A^{\varkappa}A^{\nu}A^{\rho}A^{\mu}g^{\sigma\alpha} ((\partial_{\sigma}\partial_{\alpha}\partial_{\varkappa}g_{\mu\nu})\delta g_{\rho\lambda}-(\partial_{\sigma}\partial_{\alpha}\partial{\rho}g_{\mu\nu})\delta g_{\lambda\varkappa})\\&-f^{(1)}_{D}\sqrt{-g}A^{\lambda}A^{\rho}A^{\mu}A^{\nu}g^{\sigma\alpha}
((\partial_{\sigma}\partial_{\alpha}\partial_{\lambda}A_{\mu})\delta g_{\rho\nu}-(\partial_{\sigma}\partial_{\alpha}\partial_{\lambda}A_{\rho})\delta g_{\mu\nu})+...=\\&=0+...
\label{2.4}
\end{aligned}
\end{equation*}
Now, $\frac{\delta(\sqrt{-g}\mathscr{L}_{(1)})}{\sqrt{-g}\delta A_{\sigma}}\delta A_{\sigma}$ does not contain third order derivatives as well. Indeed,
\begin{equation*}
\begin{aligned}
\delta(\sqrt{-g}\mathscr{L}_{1})&=\frac{1}{2}\sqrt{-g}\Big[f_{B}\Big(\frac{\delta(F_{;\tau}F^{;\tau})}{4}  \square F+\frac{ \square (F_{;\tau}F^{;\tau})}{4}\delta F)\Big)\\&+f_{D}((\square F)\delta D+(\square D)\delta F)\Big]+...\\&\Rightarrow\frac{1}{2}\sqrt{-g}\Big[0.5f_{B}((F_{;\tau}-F_{;\tau})\square F^{;\tau})\delta F+\\&+f_{D}(A^{\lambda}A_{\rho}A^{\mu}A_{\nu}\partial _{\tau}\partial_{\mu}\partial^{ \tau}g^{\rho\nu}-A^{\lambda}A^{\rho}A_{\mu}A_{\nu}\partial _{\tau}\partial_{\rho}\partial^{\tau}g^{\mu\nu})\delta A_{\lambda}\Big]\\&+...=0+....
\label{2.5}
\end{aligned}
\end{equation*}
Thus, equation of motion has derivatives of second order and/or lower. Summarizing, we see that the Lagrangian (\ref{L1}) leads to the second order and/or lower  field equation and energy-momentum tensor.

We now turn to the Lagrangian (\ref{L2}). Using the fact that $B=\frac{F_{;\mu}F^{;\mu}}{4}$ and $D=\frac{F^{;\mu}A_{\mu}}{2},$ we find that  
\begin{equation*}
\begin{aligned}
  &\delta(\sqrt{-g}\mathscr{L}_{2})=\frac{\sqrt{-g}}{4}\delta\Big(f^{(2)}(B, D,F)F^{;\nu}B_{;\nu}\Big)+...\Rightarrow -\frac{\sqrt{-g}}{4} \delta\Big(f^{(2)}(\square F)B\Big)\\&-\frac{\sqrt{-g}}{4} \delta\Big(f^{(2)}_{;\nu}F^{;\nu}B\Big)+...=-\frac{\sqrt{-g}}{4} \delta\Big(f^{(2)}_{;\nu}F^{;\nu}B\Big)+...\Rightarrow-\frac{\sqrt{-g}f^{(2)}_{B}}{4}\Big(-(BF^{;\nu})_{;\nu}\delta B\\
  &-(\square B)B\delta F
  +B_{;\nu}F^{;\nu}\delta B\Big)-\frac{\sqrt{-g}f^{(2)}_{D}}{4}\Big(- (F^{;\nu}B)_{;\nu}\delta D-B (\square D)\delta F+D_{;\nu}F^{;\nu}\delta B \Big)+...\\&\Rightarrow\frac{\sqrt{-g}f^{(2)}_{B}B}{8}\Big(-F^{;\tau}\square(F_{;\tau})
  +(\square F)_{;\tau}F^{;\tau}\Big)\delta F \\&~~~~~~~~~~~~~
  -\frac{\sqrt{-g}f^{(2)}_{D}A^{\lambda}F^{;\nu}}{16}\Big(F^{;\tau}F_{;\tau\nu\lambda} 
  -F^{;\tau}F_{;\lambda\nu\tau}\Big)\delta F=0+...,
\label{2.6}
\end{aligned}
\end{equation*}
where omitted terms do not contain third derivatives and  $F_{;\nu}=\nabla_{\nu}F,$ $F_{;\nu\mu}=\nabla_{\mu}\nabla_{\nu}F,$ etc., $\delta B=\frac{\delta B}{\delta g_{\rho\sigma}}\delta g_{\rho\sigma}+\frac{\delta B}{\delta A_{\mu}}\delta A_{\mu},$  etc.
Thus, all field equations have  derivatives of second order and/or lower.

The minimal extension of the Lagrangian (\ref{L3}) leads to the third order field equations. We were unable to find additional terms involving Riemann tenzor that would give rise to cancellation of the third derivatives in the field equations. Thus, we conjecture that the Lagrangians (\ref{L3}) cannot be generalized to the theory with dynamical gravity in such a way that the equations of motion remain second order. We do not consider the Lagrangian (\ref{L3}) in what follows. 

To summarize, in the case when we switch on the dynamical gravity  all field equations remain second order for two Lagrangians (\ref{L1}), (\ref{L2}).
\vspace{0.3 cm}

\section{Stable NEC-violating solution in  Minkowski space.} 
\subsection{The solution.}
Our purpose here is to figure out if there are  Lagrangians in the set (\ref{L1}), (\ref{L2}), which lead to  stable NEC-violation solutions. In  this Section we give such an example in the Minkowski background.
Let us consider the  Lagrangian  (\ref{L1}) with additional first order terms 
\begin{equation}
\mathscr{L}_{1}=qD^{2}A^{\rho}\square A_{\rho}+kB^{2}+lC^{2}+vF^{6},
\label{4.1}
\end{equation}
where $q,$ $k,$ $l$ and $v$ are free parameters and $B,$ $C,$ $D,$ $F$ are given by eqs. (\ref{1l}) -- (\ref{4l}), respectively. The specific choice of the Lagrangian functions here is such that all terms have the same transformation property
under rescaling $x^{\mu}\Rightarrow \lambda x^{\mu},$ $A_{\mu}\Rightarrow \lambda^{-1}A_{\mu},$ namely, $\mathscr{L}_{1}\Rightarrow \lambda^{-12}\mathscr{L}_{1}.$
Then there exists a non-trivial homogeneous solution of the field equation
\begin{equation}
A^{bg}_{\mu}=(\beta t^{-1},0,0,0), \;\; t>0.
\label{4.2}
\end{equation}
For this solution the field equation gives
\begin{equation}
\beta={\Big(\frac{3k+3l-5q}{v}\Big)}^{1/4}.
\label{4.3}
\end{equation}
This solution exists when 
\begin{equation}
\begin{aligned}
&3k+3l-5q>0 \;\; \mbox{and} \;\; v>0 \;\; \mbox{or} \\
&3k+3l-5q<0 \;\; \mbox{and} \;\; v<0.
\end{aligned}
\label{sol}
\end{equation}
We wil need the expression for the energy-momentum tensor for this solution: 
\begin{equation*}
\begin{aligned}
T_{\mu\nu}\big|_{g_{\rho\sigma}=\eta_{\rho\sigma};\;\;A_{\tau}=A_{\tau}^{bg}}= \frac{2\delta ( \sqrt{-g}\mathscr{L})}{\sqrt{-g}\delta g^{\mu\nu}}\big|_{g_{\rho\sigma}=\eta_{\rho\sigma};\;\;A_{\tau}=A_{\tau}^{bg}}\ .
\end{aligned}
\end{equation*}
To this end, we again consider minimal coupling to the metric, i.e., set $\square A_{\rho}=\nabla^{\mu}\nabla_{\mu}A_{\rho}$ and $D=A_{\mu;\nu}A_{\tau}A_{\lambda}g^{\mu\tau}g^{\nu\lambda},$ etc., in   
curved space-time. The Lagrangian (\ref{4.1}) can be written in the following form:
\begin{equation*}
\mathscr{L}_{1}=\frac{1}{2}f^{(1)}(D)\square F-f^{(1)}(D)A_{\tau;\sigma}A^{\tau;\sigma}+L(B,C,D,F)
\end{equation*}
where
$$f^{(1)}(D)=qD^{2},$$  $$L(B,C,D,F)=kB^2+lC^2+vF^6.$$ 
Using the fact that $\partial_{0}T^{0\rho}\big|_{A_{\mu}=A^{bg}_{\mu}}=0$ we find that $T_{00}=0$ and
\begin{equation*}
\begin{aligned}
&T_{ij}=p\delta_{ij}, \\
&p=\Big(-\frac{1}{2}\partial_{\tau}f\partial^{\tau}F+L-fA_{\tau;\sigma}A^{\tau;\sigma} \Big)\Big|_{g_{\mu\nu}=\eta_{\mu\nu};\;\;A_{\mu}=A_{\mu}^{bg}},
\label{4.4}
\end{aligned}
\end{equation*}
where $i,j=1,2,3.$
This gives 
\begin{equation*}
p=\beta^{8}t^{-12}(v\beta^{4}+k+l-9q)=\beta^{8}t^{-12}(4(k+l)-14q).
\end{equation*}
Thus, the background (\ref{4.2}) violates the NEC provided that
\begin{equation}
l+k<\frac{7q}{2}.
\label{4.5}
\end{equation}
This is possible in both cases listed in (\ref{sol}).
\subsection{Stability conditions in  Minkowski space.}
Let us consider the stability of the solution (\ref{4.2}). Having in mind Refs. \cite{12,13}, we also require subluminality of the perturbations  about it. To this end, we study somewhat more general Lagrangian
\begin{equation}
\mathscr{L}_{1}=f^{(1)}(B,D,F)A^{\rho}\square A_{\rho}+L(B,D,F,C),
\label{3.3}
\end{equation}
where $L(B,D,F)$ and $f^{(1)}(B,D,F,C)$ are arbitary functions of their arguments.
We consider homogeneous background $A^{bg}_{\mu}=(A^{bg}_{0}(t),0,0,0)$ and expand the Lagrangian (\ref{3.3}) up to the second order. In the expansion  we are only interested in coefficients of $(\delta A^{0, i} \delta A_{0}^{\ ,i} ),$ $(\delta A^{0, 0} \delta A_{0, 0} )$ and $(\delta A^{i,0} \delta A_{\ ,0}^{ i}),$ $(\delta A^{i,j} \delta A^{i,j}),$ because here we consider high momentum regime, meaning that the variation of $\delta A_{\mu}$ in space and time occurs at scales much  shorter than the time scale charactersitic of the background $A_{\mu}^{bg}(t);$ the terms $\delta A_{0,i}\delta A_{i,0}$ are not prezent. We find
\begin{equation*}
\begin{aligned}
\delta \mathscr{L}_{1}&=\mathscr{L}_{1}(A_{\mu}^{bg}+\delta A_{\mu})-\mathscr{L}_{1}(A_{\mu}^{bg}) = K_{01}(\delta A^{0, i} \delta A_{0}^{\ ,i} )+	K_{00}(\delta A^{0, 0} \delta A_{0, 0} ) \\& +K_{10}(\delta A^{i,0} \delta A_{\ ,0}^{ i})+ K_{11}(\delta A^{i,j} \delta A^{i,j} )+ (...)(\delta A^{0} \delta A_{0})+(...)(\delta A^{i} \delta A^{i}) \\& +...,
\label{3.4}
\end{aligned}
\end{equation*}
where dots denote terms with less than two derivatives, and $A \equiv A^{bg}_{0}.$ Here
\begin{equation*}
\begin{aligned}
K_{00}&= 2\dot{A}^2A(L_{BB}+L_{CC})+4L_{BC}A^4\dot{A}+ \frac{1}{2} A^4L_{DD} \\& +2A^4\dot{A}( L_{BD} +L_{CD})+2A^5\dot{A}^2\ddot{A} (f^{(1)}_{BB})+  \frac{1}{2} A^5\ddot{A}f^{(1)}_{DD}  \\&+2A^5\dot{A} \ddot{A} (f^{(1)}_{BD} )+A^3\ddot{A}( f^{(1)}_{B}) \\&-\frac{1}{2} \frac{d}{dt} (2A^3\dot{A}(f^{(1)}_{B})+A^3f^{(1)}_{D})-f^{(1)}-2A^2\dot{A}^2(f^{(1)}_{B})-2A^2\dot{A}f^{(1)}_{D} \\&-2A^2f^{(1)}_{F}+A^2(L_{C}+L_{B}),
\label{3.5}
\end{aligned}
\end{equation*}\\
\begin{equation*}
\begin{aligned}
K_{01}&=-L_{B}A^2-A^3\ddot{A}f^{(1)}_{B}+f^{(1)}-\frac{1}{2} \frac{d}{dt}(2A^3\dot{A}(f^{(1)}_{B})+A^3f^{(1)}_{D}) \\&+2A^2\dot{A}^2(f^{(1)}_{B})+2A^2\dot{A}f^{(1)}_{D}+2A^2f^{(1)}_{F},
\label{3.6}
\end{aligned}
\end{equation*}\\
\begin{equation*}
\begin{aligned}
K_{10}=f^{(1)}-L_{C}A^2,
\label{3.7}
\end{aligned}
\end{equation*}\\
\begin{equation*}
\begin{aligned}
K_{11}=-f^{(1)}.
\end{aligned}
\end{equation*}
So, the conditions of stability are
\begin{equation}
\begin{aligned}
&K_{00}>0, \ \
&K_{01}<0, \ \
&K_{10}>0, \ \
&K_{11}<0	
\end{aligned}
\label{3.8}
\end{equation}
and the condition of the absence of  superluminal perturbations  is
\begin{equation}
\begin{aligned}
&|K_{00}|>|K_{01}|,\ \
&|K_{10}|>|K_{11}|.
\label{3.9}
\end{aligned}
\end{equation}
The conditions (\ref{3.8}) and (\ref{3.9}) for the Lagrangian (\ref{4.2}) read 
\begin{equation}
\begin{aligned}
&l>3q-k,\\
&l<0,\\
&l>\frac{36q-12k}{13},\\
&v>0.
\label{3.10}
\end{aligned}
\end{equation} 
We see that the Lagrangian (\ref{4.2}) gives rise to the stable homogeneous NEC-violating solution (\ref{4.3})  when the  parameters  satisfy the relations (\ref{3.10}), (\ref{4.5}) and (\ref{sol}). In fact, all these conditions are satisfied provided that
$$v>0, \ \ q>0, \ \ l<0, \ \  3q<k<\frac{19q}{2}, \ \ \frac{36q}{13}-\frac{12k}{13}<l<\frac{7q}{2}-k.$$ 
Thus, our example shows that there are stable homogeneous solutions in our vector theories that violate the NEC.

\begin{appendices}
\section{}
\addcontentsline{toc}{section}{Appendix A}
As we discussed in section 2, we have 7 possibilities for the structure of the function $K^{\mu\nu\rho}:$\\
\begin{itemize}
\item[\RNumb{1}.] $K^{\mu\nu\rho}=L^{\mu\nu}_{\ \ \ \varkappa}(A_{\sigma}, A_{\tau;\lambda})A^{\varkappa;\rho}$
\item[\RNumb{2}.]$K^{\mu\nu\rho}=f^{\mu}(A_{\sigma}, A_{\tau;\lambda})\eta^{\nu\rho}$ 
\item[\RNumb{3}.] $K^{\mu\nu\rho}=B^{\mu\nu}(A_{\sigma}, A_{\tau;\lambda})A^{\rho}$	
\item[\RNumb{4.1}.]$K^{\mu\nu\rho}=T^{\mu}(A_{\sigma}, A_{\tau;\lambda})A^{\rho;\nu}$
\item[\RNumb{4.2}.]$K^{\mu\nu\rho}=X^{\mu\nu}_{\ \ \ \alpha}(A_{\sigma}, A_{\tau;\lambda})A^{\alpha;}_{\ \ \varkappa}A^{\rho;\varkappa}$
\item[\RNumb{4.3}.]$K^{\mu\nu\rho}=Z^{\mu\nu}_{\ \ \ \ \alpha}(A_{\sigma}, A_{\tau;\lambda})A^{\ \ \alpha}_{\varkappa;}A^{\rho;\varkappa}$
\item[\RNumb{4.4}.]$K^{\mu\nu\rho}=V^{\mu\nu}(A_{\sigma}, A_{\tau;\lambda})A_{\varkappa}A^{\rho;\varkappa}.$\\
\end{itemize}
\subsection*{Case \RNumb{1}.}
Considering option \RNumb{1}, we find that the requirement (\ref{1.9}) is equivalent to
\begin{equation}
A^{\varkappa;\rho}\frac{\partial L^{(\mu\nu)}_{\ \ \ \varkappa}}{\partial A_{\sigma;\lambda}}-
A^{\varkappa;\sigma}\frac{\partial L^{(\mu\nu)}_{\ \ \ \varkappa}}{\partial A_{\rho;\lambda}}+
L^{(\mu\nu)\sigma}\eta^{\rho\lambda}-L^{(\mu\nu)\rho}\eta^{\sigma\lambda}=0,
\label{1.10}
\end{equation}
where parenthesis denotes symmetrization.
$L^{\mu\nu\tau}$ is a monomial, so $L^{(\mu\nu)\tau}$ can be represented in the following form:
\begin{equation*}
L^{(\mu\nu)\rho}=(A_{\ \tau}^{\tau;})^{n}\widetilde{L}^{(\mu\nu)\rho},
\end{equation*}
where $n$ is non-negative integer, and $\widetilde{L}$ does not contain $A_{\ \tau}^{\tau;}.$ 
So, eq. (\ref{1.10}) reads
\begin{equation}
\eta^{\rho\lambda}\Big(-A^{\varkappa;\sigma}n(A_{\ \tau}^{\tau;})^{n-1}\widetilde{L}^{(\mu\nu)}_{\ \ \ \varkappa}+(A_{\ \tau}^{\tau;})^{n}\widetilde{L}^{(\mu\nu)\sigma} \Big)+...=0,
\label{1.11}
\end{equation}
where omitted terms do not contain the structures proportional to $\eta^{\rho\lambda}.$  We see that (\ref{1.11}) cannot be satisfied because the two terms in parenthesis  have  different powers of $A_{\ \tau}^{\tau;}.$ Thus, option \RNumb{1} does not work.\\
\subsection*{Case \RNumb{2}.}

  Considering option \RNumb{2}, we find that the requairement (\ref{1.9}) is equivalent to:
\begin{equation}
\frac{1}{2}\Big( \eta^{\rho\nu}\frac{\partial f^{\mu}}{\partial A_{\tau;\lambda}}+\eta^{\rho\mu}\frac{\partial f^{\nu}}{\partial A_{\tau;\lambda}}-\eta^{\tau\nu}\frac{\partial f^{\mu}}{\partial A_{\rho;\lambda}}-\eta^{\tau\mu}\frac{\partial f^{\nu}}{\partial A_{\rho;\lambda}}\Big)=0.
\label{1.12}
\end{equation}
We have three possibilities for function $f^{\mu}$:
\begin{enumerate}
\item[\RNumb{2}a.] $f^{\mu}=A^{\mu}h(A_{\sigma}, A_{\nu;\lambda})$
\item[\RNumb{2}b.] $f^{\mu}=A^{\mu;\varkappa}v_{\varkappa}(A_{\sigma}, A_{\nu;\lambda})$
\item[\RNumb{2}c.] $f^{\mu}=A^{\varkappa;\mu}v_{\varkappa}(A_{\sigma}, A_{\nu;\lambda})$\\
\end{enumerate}
In the case \RNumb{2}a we obtain that (\ref{1.12}) is  equivalent to
\begin{equation*}
\eta^{\nu\rho}\frac{\partial h}{\partial A_{\tau;\lambda}}=\eta^{\nu\tau}\frac{\partial h}{\partial A_{\rho;\lambda}} \ ,
\label{1.13}
\end{equation*}
which can be satisfied in the only case $h=h(A_{\sigma}),$ so that 
\begin{equation*}
f^{\mu}=A^{\mu}h(A_{\sigma}).
\end{equation*}
However the corresponding Lagrangian $\mathscr{L}=h(A_{\sigma})A^{(\mu}\eta^{\nu)\rho}A_{\rho;\mu\nu}$, does not contain  second order derivatives after  integration by parts.

In the case \RNumb{2}b we find that (\ref{1.12}) is  equivalent to
\begin{equation*}
A^{\mu;\varkappa}\eta^{\nu\rho}\frac{\partial v_{\varkappa}}{\partial A_{\tau;\lambda}}=A^{\mu;\varkappa}\eta^{\nu\tau}\frac{\partial v_{\varkappa}}{\partial A_{\rho;\lambda}} \ .
\label{1.14}
\end{equation*}
This is possible only if $v_{\varkappa}=v_{\varkappa}(A_{\sigma}).$
This leads to the following Lagrangian:
\begin{equation*}
\mathscr{L}=A_{\varkappa}(A^{\tau}A_{\tau})^{n}\eta^{\rho(\nu}A^{\mu);\varkappa}A_{\rho;\mu\nu}.
\label{1.145}
\end{equation*}
It can be reduced by integration by parts to a Lagrangian involving first derivatives only:
\begin{equation*}
\begin{aligned}
&A_{\varkappa}(A^{\tau}A_{\tau})^{n}\eta^{\rho(\nu}A^{\mu);\varkappa}A_{\rho;\mu\nu}=A_{\varkappa}(A^{\tau}A_{\tau})^{n}\frac{1}{2}(\eta^{\rho\nu}A^{\mu;\varkappa}+\eta^{\rho\mu}A^{\nu;\varkappa})A_{\rho;\mu\nu} \\&\Rightarrow-(A^{\tau}A_{\tau})^{n}A^{\varkappa}A^{\mu;}_{\ \ \ \mu\varkappa}A^{\rho;}_{\  \ \rho}+...\Rightarrow\frac{1}{2}\Big((A^{\tau}A_{\tau})^{n}A_{\varkappa}\Big)_{;\varkappa}(A^{\rho;}_{\  \ \rho})^{2}+...=0+...,
\label{1.146}
\end{aligned}
\end{equation*}
where omitted terms do not contain second derivatives, and arrows denote integration by parts.

Finally, in the case \RNumb{2}c eq. (\ref{1.12}) is  equivalent to
\begin{equation*}
v^{\rho}\eta^{\mu\tau}\eta^{\nu\lambda}=-\eta^{\mu\tau}A^{\varkappa;\nu}\frac{\partial v_{\varkappa}}{\partial A_{\rho;\lambda}}
\label{1.15}.
\end{equation*}
This equation cannot be satisfied.

Summarizing, we see that option \RNumb{2}  does not lead to  desired Lagrangians.

\subsection*{Option \RNumb{3}}
Let us consider option \RNumb{3}. It is convenient to classify the functions $B^{\mu\nu}$ according to the "origin" of the indices $\mu,$ $\nu.$ In this way we arrive at 9 possibilities (other options give the same $S^{\mu\nu\rho}$ in (\ref{1.4})):  
\begin{enumerate}
	\item[\RNumb{3}a.] $B^{\mu\nu}=h(A_{\theta}, A_{\tau;\lambda})A^{\mu}A^{\nu}$
	\item[\RNumb{3}b.] $B^{\mu\nu}=h(A_{\theta}, A_{\tau;\lambda})\eta^{\mu\nu}$
	\item[\RNumb{3}c.] $B^{\mu\nu}=h(A_{\theta}, A_{\tau;\lambda})A^{\mu;\nu}$
	\item[\RNumb{3}d.] $B^{\mu\nu}=v_{\xi}(A_{\theta}, A_{\tau;\lambda})A^{\mu;\xi}A^{\nu}$
	\item[\RNumb{3}e.] $B^{\mu\nu}=v_{\xi}(A_{\theta}, A_{\tau;\lambda})A^{\xi;\mu}A^{\nu}$
	\item[\RNumb{3}f.] $B^{\mu\nu}=L_{\xi\phi}(A_{\theta}, A_{\tau;\lambda})A^{\mu;\xi}A^{\nu;\phi}$
	\item[\RNumb{3}g.] $B^{\mu\nu}=L_{\xi\phi}(A_{\theta}, A_{\tau;\lambda})A^{\xi;\mu}A^{\phi;\nu}$
	\item[\RNumb{3}h.] $B^{\mu\nu}=L_{\xi\phi}(A_{\theta}, A_{\tau;\lambda})A^{\xi;\mu}A^{\nu;\phi}$
	\item[\RNumb{3}i.] $B^{\mu\nu}=L_{\xi\phi}(A_{\theta}, A_{\tau;\lambda})A^{\mu;\xi}A^{\phi;\nu}$
\end{enumerate}
\subsubsection*{Cases \RNumb{3}a and \RNumb{3}b.}
In  cases \RNumb{3}a, \RNumb{3}b we  obtain that the requirement (\ref{1.9}) is equivalent to
\begin{equation}
A^{\rho}\frac{\partial h}{\partial A_{\tau;\lambda}}-A^{\tau}\frac{\partial h}{\partial A_{\rho;\lambda}}=0.
\label{1.20}
\end{equation}
This is possible only if
\begin{equation*}
h=(F)^{l}(D)^{n}(B)^{k},
\label{1.22}
\end{equation*}
where $n,$ $l$ and $k$ are non-negative integers and
$$
F=A^{\rho}A_{\rho},
$$
$$
D=A^{\nu}A^{\lambda}A_{\nu; \lambda},
$$
$$
B=A^{\nu}A_{\mu}A^{\mu;\lambda}A_{\nu; \lambda}.
$$\\
Thus, this option leads to the following Lagrangians:
\begin{equation} 
\mathscr{L}_{1}=(F)^{l_{1}}(D)^{n_{1}}(B)^{k_{1}}A^{\mu}A^{\nu}A^{\rho}A_{\rho;\mu\nu},
\label{1.25}
\end{equation}
\begin{equation} 
\mathscr{L}_{2}=(F)^{l_{2}}(D)^{n_{2}}(B)^{k_{2}}\eta^{\mu\nu}A^{\rho}A_{\rho;\mu\nu},
\label{1.26}
\end{equation}
where $l_{1,2},$ $k_{1,2},$ $n_{1,2}$  are non-negative integers. We consider these Lagrangians, along with other cases, in the end of this Appendix to figure out which of them are independent.

\subsubsection*{Case  \RNumb{3}c.}
In the case (1.3) we obtain the following function $S^{\mu\nu\rho}:$
\begin{equation}
S^{\mu\nu\rho}=\frac{h}{2}(A^{\mu;\nu}+A^{\nu;\mu})A^{\rho},
\label{1.29}
\end{equation}
Using the (\ref{1.29}), we  find that the requirement (\ref{1.9}) is equivalent to the:
\begin{equation}
\eta^{\tau(\mu}\eta^{\nu)\lambda}hA^{\rho}-\eta^{\rho(\mu}\eta^{\mu)\lambda}hA^{\tau}+A^{(\mu;\nu)}\Big(\frac{\partial (hA^{\rho})}{\partial A_{\tau;\lambda}}-
\frac{\partial (hA^{\tau})}{\partial A_{\rho;\lambda}}\Big)=0.
\label{1.30}
\end{equation}
We see that (\ref{1.30}) cannot be satisfied because the first term in (\ref{1.30}) cannot be canceled out by other terms. Thus, option \RNumb{3}c does not work.

\subsubsection*{Case  \RNumb{3}d.}
In the case \RNumb{3}d we obtain the following function $S^{\mu\nu\rho}:$
\begin{equation}
S^{\mu\nu\rho}=\frac{v_{\varkappa}}{2}(A^{\mu;\varkappa}A^{\nu}+A^{\nu;\varkappa}A^{\mu})A^{\rho}.
\label{1.31}
\end{equation}
Using the (\ref{1.31}), we observe that the requarement (\ref{1.9}) is equivalent to
\begin{equation*}
\Big(\frac{\partial (A^{\rho}v_{\varkappa})}{\partial A_{\tau;\lambda}}-\frac{\partial ( A^{\tau}v_{\varkappa})}{\partial A_{\rho;\lambda}}\Big)A^{(\mu;\nu)}+A^{\rho}v^{\lambda}A^{(\nu}\eta^{\mu)\tau}-A^{\tau}v^{\lambda}A^{(\nu}\eta^{\mu)\rho}=0.
\label{1.32}
\end{equation*}
We see that (\ref{1.31}) cannot be satisfied because the third term in (\ref{1.31}) cannot be canceled out by other terms. Thus, the case \RNumb{3}d  does not lead to  desired Lagrangians.

\subsubsection*{Case  \RNumb{3}e.}
In the case \RNumb{3}e we obtain the following function $S^{\mu\nu\rho}:$
\begin{equation*}
S^{\mu\nu\rho}=\frac{v_{\varkappa}}{2}(A^{\varkappa;\mu}A^{\nu}+A^{\varkappa;\nu}A^{\mu})A^{\rho}.
\label{1.33}
\end{equation*}
Then the requirement (\ref{1.9}) is equivalent to
\begin{equation}
\begin{aligned}
&\frac{\partial (A^{\rho}v_{\varkappa})}{\partial A_{\tau;\lambda}}-\frac{\partial ( A^{\tau}v_{\varkappa})}{\partial A_{\rho;\lambda}}=0,\\
&A^{\rho}v_{\varkappa}=A_{\varkappa}v^{\rho}. 
\end{aligned}
\label{1.34}
\end{equation}
Using the second equation in (\ref{1.34}), we find that $v^{\varkappa}$ must have the following form:
\begin{equation*}
v_{\varkappa}=A_{\varkappa}h(A_{\sigma}, A_{\mu;\nu})
\label{1.341}.
\end{equation*}
From this we obtain that $h$ must obey  eq. (\ref{1.20}), so we get the Lagrangian
\begin{equation}
\mathscr{L}_{3}=(F)^{l_{3}}(D)^{n_{3}}(B)^{k_{3}}A_{\varkappa}A^{\varkappa;(\nu}A^{\mu)}A^{\rho}A_{\rho;\mu\nu},
\label{1.35}
\end{equation}
where $l_{3},$ $k_{3},$ $n_{3}$  are  non-negative integers. We consider this Lagrangian in the end of Appendix.
 
\subsubsection*{Case \RNumb{3}f.}
In the case \RNumb{3}f we obtain the following function $S^{\mu\nu\rho}:$
\begin{equation}
S^{\mu\nu\rho}=L_{(\varkappa\tau)}A^{\mu;\varkappa}A^{\nu;\sigma}A^{\rho},
\label{1.37}
\end{equation}
Using the (\ref{1.37}), we  find that the requirement (\ref{1.9}) is equivalent to
\begin{equation}
\begin{aligned}
\Big(\frac{(\partial A^{\rho}L_{(\varkappa\tau)})}{\partial A_{\sigma;\lambda}}-
\frac{\partial (A^{\sigma}L_{(\varkappa\tau)})}{\partial A_{\rho;\lambda}}\Big)A^{\mu;\varkappa}A^{\nu;\tau}+A^{\rho}(L^{(\lambda\varkappa)}\eta^{\sigma\nu}A^{\mu;}_{\ \ \ \xi}+L^{(\lambda\varkappa)}\eta^{\sigma\mu}A^{\nu;}_{\ \ \ \varkappa})\\-
A^{\sigma}(L^{(\lambda\varkappa)}\eta^{\rho\nu}A^{\mu;}_{\ \ \ \varkappa}+L^{(\lambda\varkappa)}\eta^{\rho\mu}A^{\nu;}_{\ \ \ \varkappa})=0.
\label{1.38}
\end{aligned}
\end{equation}
We see that (\ref{1.38}) cannot be satisfied because the third term in (\ref{1.38}) cannot be canceled out by other terms. Thus, the case \RNumb{3}f  does not lead to  desired Lagrangians.

\subsubsection*{Case \RNumb{3}g.}
In the case \RNumb{3}g we find the following function $S^{\mu\nu\rho}:$
\begin{equation}
S^{\mu\nu\rho}=L_{(\varkappa\tau)}A^{\varkappa;\mu}A^{\tau;\nu}A^{\rho},
\label{1.39}
\end{equation}
Using the (\ref{1.39}), we  obtain that the requirement (\ref{1.9}) is equivalent to
\begin{equation}
\begin{aligned}
&A^{\rho}L^{(\varkappa\tau)}=A^{\tau}L^{(\varkappa\rho)},\\
&\frac{\partial (A^{\rho}L^{(\varkappa\tau)}) }{A_{\sigma;\lambda}}-\frac{\partial (A^{\sigma}L^{(\varkappa\tau)}) }{A_{\rho;\lambda}}=0.
\end{aligned}
\label{1.40}
\end{equation}
Using the first equation in (\ref{1.40}), we find that $L^{\mu\nu}$ must have the following form:
\begin{equation*}
L^{\mu\nu}=A^{\mu}A^{\nu}h(A_{\theta}, A_{\xi;\tau})
\label{1.401}.
\end{equation*}\\
From this we obtain that $h$ must satisfy eq. (\ref{1.20}), and the  Lagrangian is
\begin{equation}
\mathscr{L}_{4}=(F)^{l_{4}}(D)^{n_{4}}(B)^{k_{4}}A_{\varkappa}A_{\lambda}A^{\varkappa;\mu}A^{\lambda;\nu}A^{\rho}A_{\rho;\mu\nu},
\label{1.41}
\end{equation}
where $l_{4},$ $k_{4},$ $n_{4}$  are numbers.

\subsubsection*{Case  \RNumb{3}h.}
In the case (1.8) we obtain the following function $S^{\mu\nu\rho}:$
\begin{equation} 
S^{\mu\nu\rho}=\frac{1}{2}L_{\varkappa\tau}(A^{\varkappa;\mu}A^{\nu;\tau}+A^{\varkappa;\nu}A^{\mu;\tau})A^{\rho},
\label{1.43}
\end{equation}
Using the (\ref{1.43}), we  find that the requarement (\ref{1.9}) is equivalent to
\begin{equation}
\begin{aligned}
&\frac{1}{2}(A^{\varkappa;\mu}A^{\nu;\tau}+A^{\varkappa;\nu}A^{\mu;\tau})\Big(A^{\rho}\frac{\partial (L_{\varkappa\tau})}{\partial A_{\sigma;\lambda}}-A^{\theta}\frac{\partial (L_{\varkappa\tau})}{\partial A_{\rho;\lambda}}\Big)\\&
+A^{\rho}L_{\varkappa\tau}(\eta^{\lambda(\mu}A^{\nu);\tau}\eta^{\varkappa\sigma}+A^{\varkappa;(\mu}\eta^{\nu)\sigma}\eta^{\tau\lambda})-A^{\sigma}L_{\varkappa\tau}(\eta^{\lambda(\mu}A^{\nu);\tau}\eta^{\varkappa\rho}+A^{\varkappa;(\mu}\eta^{\nu)\rho}\eta^{\tau\lambda})=0.
\label{1.44}
\end{aligned}
\end{equation}
This equation cannot be satisfied because the term $A^{\rho}L_{\varkappa\tau}A^{\varkappa;(\mu}\eta^{\nu)\sigma}\eta^{\tau\lambda}$ in (\ref{1.44}) cannot be canceled out by other terms. Thus, option  \RNumb{3}h does not work.
\subsubsection*{Case  (1.9).}
This case is similar to the previous one (1.8) and so it does not lead to desired Lagrangians.

Summarizing, we see that option \RNumb{3} leads to the four Lagrangians (\ref{1.25}), (\ref{1.26}), (\ref{1.35}), (\ref{1.41}).

\subsection*{Case \RNumb{4.1}.}
 Considering option \RNumb{4.1}, we find that the requirement (\ref{1.9}) is equivalent to 
\begin{equation}
A^{\rho;\mu}\frac{\partial T^{\nu}}{\partial A_{\tau;\lambda}}=A^{\tau;\nu}\frac{\partial T^{\mu}}{\partial A_{\rho;\lambda}}.
\label{A2}
\end{equation}
We have three possibilities for function $T^{\mu}$:
\begin{enumerate}
	\item[\RNumb{4.1}a.] $T^{\mu}=A^{\mu}h(A_{\sigma}, A_{\nu;\lambda})$
	\item[\RNumb{4.1}b.] $T^{\mu}=A^{\mu;\varkappa}v_{\varkappa}(A_{\sigma}, A_{\nu;\lambda})$
	\item[\RNumb{4.1}c.] $T^{\mu}=A^{\varkappa;\mu}v_{\varkappa}(A_{\sigma}, A_{\nu;\lambda})$
\end{enumerate}
In case \RNumb{4.1}a we obtain that (\ref{A2})  is equivalent to
\begin{equation*}
A^{\rho;\mu}A^{\nu}\frac{\partial h}{A_{\tau;\lambda}}=A^{\tau;\nu}A^{\mu}\frac{\partial h}{A_{\rho;\lambda}},
\label{A3}
\end{equation*}
which is satisfied in the only case $h=h(A_{\sigma}),$ so that
\begin{equation*}
T^{\mu}=A^{\mu}h(A_{\sigma}).
\end{equation*}
However, the corresponding Lagrangian  $\mathscr{L}=h(A_{\sigma})A^{\rho;(\nu}A^{\mu)}A_{\rho;\mu\nu}$ does not contain second order derivatives after integration by parts.\\

In the case \RNumb{4.1}b we find that (\ref{A2})  is equivalent to
\begin{equation}
A^{\rho;\mu}\Big(\eta^{\nu\tau}v^{\lambda}+A^{\nu;\varkappa}\frac{\partial v_{\varkappa}}{\partial A_{\tau;\lambda}}\Big)=A^{\tau;\nu}\Big(\eta^{\mu\rho}v^{\lambda}+A^{\mu;\varkappa}\frac{\partial v_{\varkappa}}{\partial A_{\rho;\lambda}}\Big).
\label{A5}
\end{equation}
We see that (\ref{A5}) cannot be satisfied because  the first term $A^{\rho;\mu}\eta^{\nu\tau}v^{\lambda}$ in (\ref{A5}) cannot be canceled out by other terms.

Finally, in the case \RNumb{4.1}c  we obtain that (\ref{A2})  is equivalent to
\begin{equation}
A^{\rho;\mu}\Big(\eta^{\nu\lambda}v^{\tau}+A^{\varkappa;\nu}\frac{\partial v_{\varkappa}}{\partial A_{\tau;\lambda}}\Big)=A^{\tau;\nu}\Big(\eta^{\mu\lambda}v^{\rho}+A^{\varkappa;\mu}\frac{\partial v_{\varkappa}}{\partial A_{\rho;\lambda}}\Big).
\label{A6}
\end{equation}
We see that (\ref{A6})  cannot be satisfied because  the first term $A^{\rho;\mu}\eta^{\nu\lambda}v^{\tau}$ in (\ref{A6}) cannot be canceled out by other terms.

Summarizing, we see that option \RNumb{4.1}  does not work.

\subsection*{Case \RNumb{4.2}.}
Considering option \RNumb{4.2}, we find that the requarement (\ref{1.9}) is equivalent to
\begin{equation}
L^{(\mu\nu)\sigma}A^{\rho;\lambda}+A^{\tau;}_{\ \ \varkappa}A^{\rho;\varkappa}\frac{\partial L^{(\mu\nu)}_{\;\;\; \ \ \tau}}{\partial A_{\sigma;\lambda}}-L^{(\mu\nu)\rho}A^{\sigma;\lambda}-A^{\tau;}_{\ \ \varkappa}A^{\sigma;\varkappa}\frac{\partial L^{(\mu\nu)}_{\;\;\;\ \ \ \tau}}{\partial A_{\rho;\lambda}}=0
\label{1.16}.
\end{equation}
$L^{\mu\nu\sigma}$ is a monomial, so $L^{(\mu\nu)\sigma}$ can be represented in the following form:
\begin{equation*}
L^{(\mu\nu)\sigma}=(A^{\varkappa;\tau}A_{\varkappa;\tau})^{n}\widetilde{L}^{(\mu\nu)\sigma}
\label{1.17},  
\end{equation*}
where $n$ is a natural number, and $\widetilde{L}^{(\mu\nu)\sigma}$ does not contain $(A^{\rho;\tau}A_{\rho;\tau}).$ 
So, eq (\ref{1.16}) reads
\begin{equation}
((A^{\varkappa;\tau}A_{\varkappa;\tau})^{n}\widetilde{L}^{(\mu\nu)\sigma}-2nA^{\tau;}_{\ \ \varkappa}A^{\sigma;\varkappa}\widetilde{L}^{(\mu\nu)}_{\;\;\;\ \ \ \tau}(A^{\varkappa;\tau}A_{\varkappa;\tau})^{n-1})A^{\rho;\lambda}+...=0
\label{1.18},
\end{equation}
where omitted terms do not contain the structures proportional to $A^{\rho;\lambda}.$ 
We see that  (\ref{1.18}) cannot be satisfied because the two terms in parenthesis  have  different powers of  $(A^{\varkappa;\tau}A_{\varkappa;\tau}).$ Thus, option \RNumb{4.2} does not work.

\subsection*{Case \RNumb{4.3}.}
Considering option \RNumb{4.3}, we find that the requarement (\ref{1.9}) is equivalent to
\begin{equation}
A^{\rho;\tau}L^{(\mu\nu)\lambda}+A^{\rho;\varkappa}A_{\varkappa;\sigma}\frac{\partial L^{(\mu\nu)\sigma}}{\partial A_{\tau;\lambda}}-A^{\tau;\rho}L^{(\mu\nu)\lambda}-A^{\tau;\varkappa}A_{\varkappa;\sigma}\frac{\partial L^{(\mu\nu)\sigma}}{\partial A_{\rho;\lambda}}=0.
\label{1.19}
\end{equation}
We see that (\ref{1.19}) cannot be satisfied because the first term in (\ref{1.19}) cannot be canceled out by others terms in (\ref{1.19}).Thus, option \RNumb{4.3} does not lead to  desired Lagrangians.\\

\subsection*{Option \RNumb{4.4}.}
We now consider option \RNumb{4.4}. It is convenient to classify the functions $Z^{\mu\nu}$ according to the "origin" of the indices $\mu,$ $\nu.$ In this way we arrive at 9 possibilities (other options give the same $S^{\mu\nu\rho}$ in (\ref{1.4})):  
\begin{enumerate}
\item[\RNumb{4.4}a.] $Z^{\mu\nu}=h(A_{\theta}, A_{\tau;\lambda})A^{\mu}A^{\nu}$
\item[\RNumb{4.4}b.] $Z^{\mu\nu}=h(A_{\theta}, A_{\tau;\lambda})\eta^{\mu\nu}$
\item[\RNumb{4.4}c.] $Z^{\mu\nu}=h(A_{\theta}, A_{\tau;\lambda})A^{\mu;\nu}$
\item[\RNumb{4.4}d.] $Z^{\mu\nu}=v_{\xi}(A_{\theta}, A_{\tau;\lambda})A^{\mu;\xi}A^{\nu}$
\item[\RNumb{4.4}e.] $Z^{\mu\nu}=v_{\xi}(A_{\theta}, A_{\tau;\lambda})A^{\xi;\mu}A^{\nu}$
\item[\RNumb{4.4}f.] $Z^{\mu\nu}=L_{\xi\phi}(A_{\theta}, A_{\tau;\lambda})A^{\mu;\xi}A^{\nu;\phi}$
\item[\RNumb{4.4}g.] $Z^{\mu\nu}=L_{\xi\phi}(A_{\theta}, A_{\tau;\lambda})A^{\xi;\mu}A^{\phi;\nu}$
\item[\RNumb{4.4}h.] $Z^{\mu\nu}=L_{\xi\phi}(A_{\theta}, A_{\tau;\lambda})A^{\xi;\mu}A^{\nu;\phi}$
\item[\RNumb{4.4}i.] $Z^{\mu\nu}=L_{\xi\phi}(A_{\theta}, A_{\tau;\lambda})A^{\mu;\xi}A^{\phi;\nu}$
\end{enumerate}
\subsubsection*{Cases \RNumb{4.4}a and \RNumb{4.4}b.}
In  cases \RNumb{4.4}a, \RNumb{4.4}b we  find that the requirement (\ref{1.9}) is equivalent to
\begin{equation*}
A^{\rho;\varkappa}A_{\varkappa}\frac{\partial h}{\partial A_{\tau;\lambda}}-A^{\tau;\varkappa}A_{\varkappa}\frac{\partial h}{\partial A_{\rho;\lambda}}=0,
\label{1.23}
\end{equation*}
which can be satisfied in the only case $h=(F)^{l}(A^{\mu;\tau}A_{\tau}A^{\rho}A_{\mu;\rho})^{n},$ so that  we have the following Lagrangians:
\begin{equation}
\mathscr{L}_{5}=(F)^{l_{3}}(C)^{n_{5}}A^{\mu}A^{\nu}A^{\rho;\varkappa}A_{\varkappa}A_{\rho;\mu\nu},
\label{1.27}
\end{equation}
\begin{equation}
\mathscr{L}_{6}=(F)^{l_{6}}(C)^{n_{6}}\eta^{\mu\nu}A^{\rho;\varkappa}A_{\varkappa}A_{\rho;\mu\nu},
\label{1.28}
\end{equation}
where $l_{5,6},$  $n_{5,6}$  are non-negative integers, and 
$$C=A^{\mu;\tau}A_{\tau}A^{\rho}A_{\mu;\rho}.$$
We discuss Lagrangians (\ref{1.27}), (\ref{1.28}) in the end of this Appendix.

\subsubsection*{Case  \RNumb{4.4}c.}
In the case \RNumb{4.4}c we obtain the following function $S^{\mu\nu\rho}:$
\begin{equation}
S^{\mu\nu\rho}=\frac{h}{2}(A^{\mu;\nu}+A^{\nu;\mu})A^{\rho}.
\label{1.029}
\end{equation}
Using the (\ref{1.029}), we  find that the requirement (\ref{1.9}) is equivalent to
\begin{equation}
\eta^{\tau(\mu}\eta^{\nu)\lambda}hA^{\rho;\varkappa}A_{\varkappa}-\eta^{\rho(\mu}\eta^{\mu)\lambda}hA^{\tau;\varkappa}A_{\varkappa}+A^{(\mu;\nu)}\Big(\frac{\partial (hA^{\rho;\varkappa}A_{\varkappa})}{\partial A_{\tau;\lambda}}-
\frac{\partial (hA^{\tau;\varkappa}A_{\varkappa})}{\partial A_{\rho;\lambda}}\Big)=0.
\label{1.030}
\end{equation}
We see that (\ref{1.030}) cannot be satisfied because the first term in (\ref{1.030}) cannot be canceled out by other terms. Thus, option \RNumb{4.4}c does not work.

\subsubsection*{Case  \RNumb{4.4}d.}
In the case \RNumb{4.4}d we obtain the following function $S^{\mu\nu\rho}:$
\begin{equation}
S^{\mu\nu\rho}=\frac{v_{\varkappa}}{2}(A^{\mu;\varkappa}A^{\nu}+A^{\nu;\varkappa}A^{\mu})A^{\rho},
\label{1.031}
\end{equation}
Using the (\ref{1.031}), we  find that the requirement (\ref{1.9}) is equivalent to
\begin{equation}
\Big(\frac{\partial (A^{\rho;\sigma}A_{\sigma}v_{\varkappa})}{\partial A_{\tau;\lambda}}-\frac{\partial ( A^{\tau;\sigma}A_{\sigma}v_{\varkappa})}{\partial A_{\rho;\lambda}}\Big)A^{(\mu;\nu)}+A^{\rho;\sigma}A_{\sigma}v^{\lambda}A^{(\nu}\eta^{\mu)\tau}-A^{\tau;\sigma}A_{\sigma}v^{\lambda}A^{(\nu}\eta^{\mu)\rho}=0.
\label{1.032}
\end{equation}
We see that (\ref{1.032}) cannot be satisfied because the third term in (\ref{1.032}) cannot be canceled out by other terms. Thus, option  \RNumb{4.4}d  does not lead to  desired Lagrangians.

\subsubsection*{Case  \RNumb{4.4}e.}
In the case \RNumb{4.4}e we find the following function $S^{\mu\nu\rho}:$
\begin{equation*}
S^{\mu\nu\rho}=\frac{v_{\varkappa}}{2}(A^{\varkappa;\mu}A^{\nu}+A^{\varkappa;\nu}A^{\mu})A^{\rho;\sigma}A_{\sigma}.
\label{1.033}
\end{equation*}
Then the requirement (\ref{1.9}) is equivalent to
\begin{equation}
\begin{aligned}
&\frac{\partial (A^{\rho;\sigma}A_{\sigma}v_{\varkappa})}{\partial A_{\tau;\lambda}}-\frac{\partial ( A^{\tau;\sigma}A_{\sigma}v_{\varkappa})}{\partial A_{\rho;\lambda}}=0,\\
&A^{\rho;\sigma}A_{\sigma}v_{\varkappa}=A_{\varkappa;\sigma}A_{\sigma}v^{\rho}. 
\end{aligned}
\label{1.034}.
\end{equation}
Using the second equation in (\ref{1.034}), we find that $v_{\varkappa}$ must have the following form:
\begin{equation*}
v_{\varkappa}=A_{\varkappa;\sigma}A^{\sigma}h(A_{\tau}, A_{\mu;\nu}).
\label{1.035}
\end{equation*}
So, eq. (\ref{1.034}) reads
\begin{equation}
A_{\varkappa}^{\ \ ;(\nu}A^{\mu)}\Big(A^{\rho;\tau}A_{\tau}A^{\varkappa;\tau}A_{\tau}\frac{\partial h}{\partial A_{\sigma;\lambda}}+A^{\rho;\tau}A^{\lambda}hA_{\tau}\eta^{\varkappa\sigma}-
A^{\sigma;\tau}A_{\tau}A^{\varkappa;\tau}A_{\tau}\frac{\partial h}{\partial A_{\rho;\lambda}}-A^{\sigma;\tau}A^{\lambda}hA_{\tau}\eta^{\varkappa\rho}\Big)=0.
\label{1.036}
\end{equation}
This equation cannot be satisfied, because the second term in (\ref{1.036}) cannot be canceled out by other terms. Thus, option \RNumb{4.4}e does not work.

\subsubsection*{Case  \RNumb{4.4}f.}
In the case \RNumb{4.4}f we find that (\ref{1.9}) is equivalent to
\begin{equation}
\begin{aligned}
\Big(\frac{\partial( A^{\rho;\sigma}A_{\sigma}L_{(\varkappa\tau)})}{\partial A_{\alpha;\lambda}}-
\frac{\partial (A^{\alpha;\sigma}A_{\sigma}L_{(\varkappa\tau)})}{\partial A_{\rho;\lambda}}\Big)A^{\mu;\varkappa}A^{\nu;\tau}+A^{\rho;\sigma}A_{\sigma}(L^{(\lambda\varkappa)}\eta^{\alpha\nu}A^{\mu;}_{\ \ \ \varkappa}+L^{(\lambda\varkappa)}\eta^{\alpha\mu}A^{\nu;}_{\ \ \ \varkappa})\\-
A^{\alpha;\sigma}A_{\sigma}(L^{(\lambda\varkappa)}\eta^{\rho\nu}A^{\mu;}_{\ \ \ \varkappa}+L^{(\lambda\varkappa)}\eta^{\rho\mu}A^{\nu;}_{\ \ \ \varkappa})=0.
\label{1.038}
\end{aligned}
\end{equation}
We see that (\ref{1.038}) cannot be satisfied because the third term in eq. (\ref{1.038}) cannot be canceled out by other terms. Thus, this option  does not lead to  desired Lagrangians.

\subsubsection*{Case  \RNumb{4.4}g.}
In the case \RNumb{4.4}g we obtain that (\ref{1.9}) is equivalent to
\begin{equation}
\begin{aligned}
&f^{\rho}L^{(\varkappa\tau)}=f^{\tau}L^{(\varkappa\rho)},\\
&\frac{\partial (f^{\rho}L^{(\varkappa\tau)}) }{A_{\sigma;\lambda}}-\frac{\partial (f^{\sigma}L^{(\varkappa\tau)}) }{A_{\rho;\lambda}}=0,
\end{aligned}
\label{1.040}
\end{equation}
where $f^{\rho}=A^{\rho;\mu}A_{\mu}.$
This is possible only if $L^{\mu\nu}=f^{\mu}f^{\nu}h(A_{\sigma}, A_{\varkappa;\tau}).$
From this we find that (\ref{1.040}) is equivalent to
\begin{equation}
\begin{aligned}
&A^{\rho;\tau}A_{\tau}A^{\varkappa;\sigma}A_{\sigma}A^{\alpha;\mu}A_{\mu}\Big(\frac{\partial h}{A_{\nu;\lambda}}\Big)-A^{\nu;\tau}A_{\tau}A^{\varkappa;\sigma}A_{\sigma}A^{\alpha;\mu}A_{\mu}\Big(\frac{\partial h}{A_{\rho;\lambda}}\Big)\\&
+h\Big(A^{\alpha;\mu}A_{\mu}A^{\rho;\tau}A_{\tau}\eta^{\varkappa\nu}A^{\lambda}-A^{\alpha;\mu}A_{\mu}A^{\nu;\tau}A_{\tau}\eta^{\varkappa\rho}A^{\lambda}+ A^{\varkappa;\mu}A_{\mu}A^{\rho;\tau}A_{\tau}\eta^{\alpha\nu}A^{\lambda}\\&-A^{\varkappa;\mu}A_{\mu}A^{\nu;\tau}A_{\tau}\eta^{\alpha\rho}A^{\lambda}    \Big)=0.
\label{1.42}
\end{aligned}
\end{equation}
We see that (\ref{1.42}) cannot be satisfied because the third term in eq. (\ref{1.42}) cannot be canceled out by other terms. Thus, option \RNumb{4.4}g does not work.

\subsubsection*{Case  \RNumb{4.4}h.}
In the case \RNumb{4.4}h we obtain that (\ref{1.9}) is equivalent to
\begin{equation}
\begin{aligned}
&\frac{1}{2}(A^{\varkappa;\mu}A^{\nu;\tau}+A^{\varkappa;\nu}A^{\mu;\tau})\Big(f^{\rho}\frac{\partial (L_{\varkappa\tau})}{\partial A_{\alpha;\lambda}}-f^{\alpha}\frac{\partial (L_{\varkappa\tau})}{\partial A_{\rho;\lambda}}\Big)\\&
+f^{\rho}L_{\varkappa\tau}(\eta^{\lambda(\mu}A^{\nu);\tau}\eta^{\varkappa\alpha}+A^{\varkappa;(\mu}\eta^{\nu)\alpha}\eta^{\tau\lambda})-f^{\alpha}L_{\varkappa\tau}(\eta^{\lambda(\mu}A^{\nu);\tau}\eta^{\varkappa\rho}+A^{\varkappa;(\mu}\eta^{\nu)\rho}\eta^{\tau\lambda})=0,
\label{1.044}
\end{aligned}
\end{equation}
where $f^{\rho}=A^{\rho;\sigma}A_{\sigma}.$ We see that (\ref{1.044}) cannot be satisfied because the  term $A^{\varkappa;(\mu}\eta^{\nu)\alpha}\eta^{\tau\lambda}$ in (\ref{1.044}) cannot be canceled out by other terms. Thus, this option  does not lead to desired Lagrangians.

\subsubsection*{Case  \RNumb{4.4}i.}
This case is similar to the previous one \RNumb{4.4}i and so it does not lead to desired Lagrangians.

\subsection*{Independent Lagrangians.}
To summarize, we have arrived at the six Lagrangians (\ref{1.25}), (\ref{1.26}), (\ref{1.35}), (\ref{1.41}), (\ref{1.27}), (\ref{1.28}). We write them again for references:
\begin{equation} 
\mathscr{L}_{1}=(F)^{l_{1}}(D)^{n_{1}}(B)^{k_{1}}A^{\mu}A^{\nu}A^{\rho}A_{\rho;\mu\nu},
\label{1}
\end{equation}
\begin{equation} 
\mathscr{L}_{2}=(F)^{l_{2}}(D)^{n_{2}}(B)^{k_{2}}\eta^{\mu\nu}A^{\rho}A_{\rho;\mu\nu},
\label{2}
\end{equation}
\begin{equation}
\mathscr{L}_{3}=(F)^{l_{3}}(D)^{n_{3}}(B)^{k_{3}}A_{\varkappa}A^{\varkappa;(\nu}A^{\mu)}A^{\rho}A_{\rho;\mu\nu},
\label{3}
\end{equation}
\begin{equation}
\mathscr{L}_{4}=(F)^{l_{4}}(D)^{n_{4}}(B)^{k_{4}}A_{\varkappa}A_{\sigma}A^{\varkappa;\mu}A^{\sigma;\nu}A^{\rho}A_{\rho;\mu\nu},
\label{4}
\end{equation}
\begin{equation}
\mathscr{L}_{5}=(F)^{l_{5}}(C)^{n_{5}}A^{\mu}A^{\nu}A^{\rho;\varkappa}A_{\varkappa}A_{\rho;\mu\nu},
\label{5}
\end{equation}
\begin{equation}
\mathscr{L}_{6}=(F)^{l_{6}}(C)^{n_{6}}\eta^{\mu\nu}A^{\rho;\varkappa}A_{\varkappa}A_{\rho;\mu\nu},
\label{6}
\end{equation}
Upon integration by parts some of these Lagrangians are reduced to the Lagrangian containing the first derivatives only. Our purpose here to figure out which of these Lagrangians are independent modulo first-order Lagrangians.

The Lagrangian (\ref{5}) can be reduced by integration by parts to a Lagrangian involving first
derivatives only:
\begin{equation*}
\begin{aligned}
&\mathscr{L}_{5}=(F)^{l_{5}}(C)^{n_{5}}A^{\mu}A^{\nu}A^{\rho;\varkappa}A_{\varkappa}A_{\rho;\mu\nu} \\
&=\frac{1}{2}(F)^{l_{5}}(C)^{n_{5}}A^{\nu}C_{;\nu}+...=\frac{1}{2(n_{5}+1)}(F)^{l_{5}}((C)^{n_{5}+1})_{;\nu}A^{\nu}+...\\
&\Rightarrow-\frac{1}{2(n_{5}+1)}(F)^{l_{5}}(C)^{n_{5}+1}A^{\nu;}_{\ \ \nu}+...=0+... ,
\end{aligned}
\end{equation*}
where, as before,  omitted terms do not contain second derivatives and arrow denotes integration by parts.

Upon integration by parts and adding terms containing first derivatives only, the remaining Lagrangians (\ref{1}) -- (\ref{4}), (\ref{6}) can be expressed throught three Lagrangians (\ref{2}), (\ref{4}), (\ref{6}). Indeed, the Lagrangian (\ref{1}) can be expressed through the Lagrangian (\ref{3}):
\begin{equation*}
\begin{aligned} 
&\mathscr{L}_{1}=(F)^{l_{1}}(D)^{n_{1}}(B)^{k_{1}}A^{\mu}A^{\nu}A^{\rho}A_{\rho;\mu\nu}+...=\frac{1}{2}(F)^{l_{1}}(D)^{n_{1}}(B)^{k_{1}}A^{\nu}D_{;\nu}+...\\&\Rightarrow-\frac{k_{1}}{2(n_{1}+1)}( (F)^{l_{1}}(D)^{n_{1}+1}(B)^{k_{1}-1} )F^{;\nu}D_{;\nu}+...\\&=-\frac{k_{1}}{(n_{1}+1)}(F)^{l_{1}}(D)^{n_{1}+1}(B)^{k_{1}-1}A_{\varkappa}A^{\varkappa;(\nu}A^{\mu)}A^{\rho}A_{\rho;\mu\nu}+...
\label{01}
\end{aligned}
\end{equation*}
The Lagrangian (\ref{3}) can in turn be expressed throught two Lagrangians (\ref{2}) and (\ref{4}):
\begin{equation*}
\begin{aligned}
  &\mathscr{L}_{3}=(F)^{l_{3}}(D)^{n_{3}}(B)^{k_{3}}A_{\varkappa}A^{\varkappa;(\nu}A^{\mu)}A^{\rho}A_{\rho;\mu\nu}+...=\frac{1}{2}(F)^{l_{3}}(D)^{n_{3}}(B)^{k_{3}}D_{;\mu}F^{;\mu}+...
  \\&\Rightarrow-\frac{1}{2(n_{3}+1)} (F)^{l_{3}}(D)^{n_{3}+1}(B)^{k_{3}} \square F -  \frac{k_{3}}{2(n_{3}+1)}(F)^{l_{3}}(D)^{n_{3}+1}(B)^{k_{3}-1}B_{;\nu}F^{;\nu} +...
  \\
&=-\frac{1}{(n_{3}+1)} (F)^{l_{3}}(D)^{n_{3}+1}(B)^{k_{3}}\eta^{\mu\nu}A^{\rho}A_{\rho;\mu\nu}-\\& -\frac{2k_{3}}{(n_{3}+1)}(F)^{l_{3}}(D)^{n_{3}+1}(B)^{k_{3}-1}A_{\varkappa}A_{\tau}A^{\varkappa;\mu}A^{\tau;\nu}A^{\rho}A_{\rho;\mu\nu} +...
\label{02}
\end{aligned}
\end{equation*}

There are four special cases in which the remaining Lagrangians (\ref{2}), (\ref{4}) and (\ref{6}) are, in fact, first order or are not independent. One is the Lagrangian (\ref{2}) with $n_{2}=0,1$ and $k_{2}=0:$
\begin{equation*}
\begin{aligned}
&F^{l_{2}}D\eta^{\mu\nu}A^{\rho}A_{\rho;\mu\nu}+...=\frac{1}{2}F^{l_{2}}D\square F+...\Rightarrow-\frac{1}{2}F^{l_{2}}D_{;\nu}F^{;\nu}+...\\
&=-\frac{1}{4}F^{l_{2}}A^{\theta}F_{;\theta\nu}F^{;\nu}+...=-\frac{1}{8}F^{l_{2}}A^{\lambda}(F_{;\nu}F^{;\nu})_{;\lambda}+...\\&\Rightarrow\frac{1}{8}(F^{l_{2}}A^{\lambda})_{;\lambda}F_{;\nu}F^{;\nu}+...=0+...
\end{aligned}
\end{equation*}
Another is the Lagrangian (\ref{4}) with $n_{4}=0$ and $k{4=0}$, which is effectively first order.\\
The third special case is the Lagrangian (\ref{2}) with $n_{2}=1$ and arbitrary $k_{2},$ which can be expressed throught the Lagrangian (\ref{4}):
\begin{equation*}
\begin{aligned}
&\mathscr{L}_{2}=(F)^{l_{2}}(B)^{k_{2}}D\eta^{\mu\nu}A^{\rho}A_{\rho;\mu\nu}=\frac{1}{2}(F)^{l_{2}}(B)^{k_{2}}D \square{F}+...\\&\Rightarrow-\frac{1}{2}(F)^{l_{2}}\big((B)^{k_{2}}D\big)_{;\nu} F^{;\nu}+...=-\frac{1}{2}(F)^{l_{2}}(B)^{k_{2}}D_{;\nu} F^{;\nu}-\frac{k_{2}}{2}(F)^{l_{2}}(B)^{k_{2}-1}D B_{;\nu}F^{;\nu}+...\\&
= -\frac{1}{2}(F)^{l_{2}}(B)^{k_{2}}B_{;\nu}A ^{\nu}                              -2k_{2}(F)^{l_{2}}(B)^{k_{2}-1}DA_{\varkappa}A_{\tau}A^{\varkappa;\mu}A^{\tau;\nu}A^{\rho}A_{\rho;\mu\nu}+...\\&\Rightarrow-2k_{2}(F)^{l_{2}}(B)^{k_{2}-1}DA_{\varkappa}A_{\tau}A^{\varkappa;\mu}A^{\tau;\nu}A^{\rho}A_{\rho;\mu\nu}+...,
\label{03}
\end{aligned}
\end{equation*}
Finally, the Lagrangian (\ref{6}) is effectively first order for $n_{6}=0.$

This completes the analysis leading to the result quoted in the end of Section 2, eqs. (\ref{1.96}) -- (\ref{f}).  

\end{appendices}

\end{document}